\documentclass[proceedings,/home/valle/Stylefiles/showkeys]{JHEP} 
\input epsf
\usepackage{psfig} 
\usepackage{epsfig} 
\def\ifmath#1{\relax\ifmmode #1\else $#1$\fi}
\def\half{\ifmath{{\textstyle{1 \over 2}}}}

\def\3quarter{{\textstyle{3 \over 4}}}

\def\lf{\leaders\hbox to 1em{\hss.\hss}\hfill}
\def\e6{$E(6)$}
\def\10{$SO(10)$}
\def\21{$SU(2) \otimes U(1) $}
\def\lr{$SU(2)_L \otimes SU(2)_R \otimes U(1)$}
\def\422{$SU(4) \otimes SU(2) \otimes SU(2)$}
\def\321{$SU(3) \otimes SU(2) \otimes U(1)$}
\def\ne{\hbox{$\nu_e$ }}
\def\nm{\hbox{$\nu_\mu$ }}
\def\nt{\hbox{$\nu_\tau$ }}
\def\ns{\hbox{$\nu_{s}$ }}

\def\O{\hbox{$\cal O$ }}

\def\mnt{\hbox{$m_{\nu_\tau}$ }}

\def\neus{\hbox{neutrinos }}

\def\neu{\hbox{neutrino }}

\def\eq#1{{eq. (\ref{#1})}}

\def\fig#1{{Fig. (\ref{#1})}}

\let\vev\VEV

\def\lsim{\raise0.3ex\hbox{$\;<$\kern-0.75em\raise-1.1ex\hbox{$\sim\;$}}}
\def\gsim{\raise0.3ex\hbox{$\;>$\kern-0.75em\raise-1.1ex\hbox{$\sim\;$}}}
\def\half{{1\over 2}}
\def\beq{\begin{equation}}
\def\eeq{\end{equation}}
\def\bef{\begin{figure}}
\def\eef{\end{figure}}
\def\bet{\begin{table}}
\def\eet{\end{table}}
\def\bea{\begin{eqnarray}}
\def\ba{\begin{array}}
\def\ea{\end{array}}
\def\bi{\begin{itemize}}
\def\ei{\end{itemize}}
\def\ben{\begin{enumerate}}
\def\een{\end{enumerate}}

\def\eea{\end{eqnarray}}
\def\apj#1#2#3{          {\it Astrophys. J. }{\bf #1} (19#2) #3}

\def\aa#1#2#3{          {\it Astron. \& Astrophys.  }{\bf #1} (19#2) #3}

\def\ib#1#2#3{           {\it ibid. }{\bf #1} (19#2) #3}
\def\nat#1#2#3{          {\it Nature }{\bf #1} (19#2) #3}
\def\nps#1#2#3{        {\it Nucl. Phys. B (Proc. Suppl.) }{\bf #1} (19#2) #3} 
\def\np#1#2#3{           {\it Nucl. Phys. }{\bf #1} (19#2) #3}
\def\pl#1#2#3{           {\it Phys. Lett. }{\bf #1} (19#2) #3}
\def\pr#1#2#3{           {\it Phys. Rev. }{\bf #1} (19#2) #3}
\def\prep#1#2#3{         {\it Phys. Rep. }{\bf #1} (19#2) #3}
\def\prl#1#2#3{          {\it Phys. Rev. Lett. }{\bf #1} (19#2) #3}

\def\zp#1#2#3{           {\it Zeit. fur Physik }{\bf #1} (19#2) #3}

\def\n.c.#1#2#3{         {\it Nuovo Cim. }{\bf #1} (19#2) #3}
\def\r.n.c.#1#2#3{       {\it Riv. del Nuovo Cim. }{\bf #1} (19#2) #3}

\def\mpl#1#2#3{          {\it Mod. Phys. Lett. }{\bf #1} (19#2) #3}

\def\ppnp#1#2#3{           {\it Prog. Part. Nucl. Phys. }{\bf #1} (19#2) #3}

\def\pc{private communication}

\def\ip{in preparation}
\def\bne{\hbox{$\bar\nu_e$ }}  
\def\bnm{\hbox{$\bar\nu_\mu$ }}  
 
\conference{ Corfu Summer Institute on Particle Physics, 1998}

\title{Neutrino Mass: theory, data and interpretation }

\author{ J. W. F. Valle\\
Instituto de F\'{\i}sica Corpuscular -- C.S.I.C. -- Univ. de
Val\`encia, Departamento de F\'{\i}sica Te\`orica \\ 46100
Burjassot, Val\`encia, Spain \\ E-mail: \email{http://neutrinos.uv.es }}

\vspace*{0.9cm} 

\abstract{
In these two lectures I describe first the theory of neutrino mass and
then discuss the implications of recent data (including 708--day data
Super--Kamiokande data) which strongly indicate the need for neutrino
conversions to account for the solar and atmospheric neutrino
observations.  I also mention the LSND data, which provides an
intriguing hint.  The simplest ways to reconcile all these data in
terms of neutrino oscillations invoke a light sterile neutrino in
addition to the three active ones.  Out of the four neutrinos, two are
maximally-mixed and lie at the LSND scale, while the others are at the
solar mass scale.  These schemes can be distinguished at
neutral-current-sensitive solar \& atmospheric neutrino experiments. I
discuss the simplest theoretical scenarios, where the lightness of the
sterile neutrino, the nearly maximal atmospheric neutrino mixing, and
the generation of $\Delta {m^2}_\odot$ \& $\Delta {m^2}_{atm}$ all
follow naturally from the assumed lepton-number symmetry and its
breaking.  Although the most likely interpretation of the present data
is in terms of neutrino-mass-induced oscillations, one still has room
for alternative explanations, such as flavour changing neutrino
interactions, with no need for neutrino mass or mixing. Such flavour
violating transitions arise in theories with strictly massless
neutrinos, and may lead to other sizeable flavour non-conservation
effects, such as $\mu \to e +
\gamma$, $\mu-e$ conversion in nuclei, unaccompanied by neutrino-less
double beta decay. }

\begin{document} 

\section{Introduction}
\vskip .1cm

Since the early geochemical experiments of Davis and collaborators,
underground experiments have by now provided solid evidence for the
solar and the atmospheric neutrino problems, the two milestones in the
search for physics beyond the Standard Model (SM)
~\cite{solarexp,atmexp,sk535a,sk300,sk504s,Smy:1999tt,Inoue99venice}.
Of particular importance has been the recent confirmation by the
Super-Kamiokande collaboration \cite{sk535a} of the atmospheric
neutrino zenith angle dependent deficit, which has marked a turning
point in our understanding of neutrinos, providing a strong evidence
for \nm conversions.  In addition to the neutrino data from
underground experiments there is also some indication for neutrino
oscillations from the LSND experiment~\cite{LSND,Louis:1998qf}.

Neutrino conversions are naturally expected to take place if neutrinos
are massive, as expected in most extensions of the Standard
Model~\cite{fae}.  The preferred theoretical origin of neutrino mass
is lepton number violation, which typically leads also to lepton
flavour violating transitions such as neutrino-less double beta decay,
so far unobserved.  However, lepton flavour violating transitions may
arise without neutrino masses ~\cite{BER,3E} in models with
extra heavy leptons \cite{WYLER,SST} and in supergravity
~\cite{Hall:1986dx}. Indeed the atmospheric neutrino anomaly can be
explained in terms of flavour changing neutrino interactions, with no
need for neutrino mass or mixing~\cite{Gonzalez-Garcia:1998hj}.
Whether or not this mechanism will resist the test of time it will
still remain as one of the ingredients of the final solution, at the
moment not required by the data.  A possible signature of theories
leading to FC interactions would be the existence of sizeable flavour
non-conservation effects, such as $\mu \to e + \gamma$, $\mu-e$
conversion in nuclei, unaccompanied by neutrino-less double beta
decay. In contrast to the intimate relationship between the latter and
the non-zero Majorana mass of neutrinos due to the Black-Box
theorem~\cite{Schechter:1982bd} there is no fundamental link between
lepton flavour violation and neutrino mass.  Barring such exotic
mechanisms reconciling the LSND (and possibly Hot Dark Matter, see
below) together with the data on solar and atmospheric neutrinos
requires {\sl three mass scales}.  The simplest way is to invoke the
existence of a light sterile neutrino~\cite{ptv92,pv93,cm93}. Out of
the four neutrinos, two of them lie at the solar neutrino scale and
the other two maximally-mixed neutrinos are at the HDM/LSND scale. The
prototype models proposed in~\cite{ptv92,pv93} enlarge the \21 Higgs
sector in such a way that neutrinos acquire mass radiatively, without
unification nor seesaw. The LSND scale arises at one-loop, while the
solar and atmospheric scales come in at the two-loop level, thus
accounting for the hierarchy. The lightness of the sterile neutrino,
the nearly maximal atmospheric neutrino mixing, and the generation of
the solar and atmospheric neutrino scales all result naturally from
the assumed lepton-number symmetry and its breaking.  Either \ne- \nt
conversions explain the solar data with \nm- \ns oscillations
accounting for the atmospheric deficit~\cite{ptv92}, or else the
r\^oles of \nt and \ns are reversed ~\cite{pv93}. These two basic
schemes have distinct implications at future solar \& atmospheric
neutrino experiments with good sensitivity to neutral current neutrino
interactions. Cosmology can also place restrictions on these
four-neutrino schemes.

\section{Mechanisms for Neutrino Mass}
\vskip .1cm

{\sl Why are neutrino masses so small compared to those of the charged
fermions}? Because of the fact that neutrinos, being the only
electrically neutral elementary fermions should most likely be
Majorana, the most fundamental fermion. In this case the suppression
of their mass could be associated to the breaking of lepton number
symmetry at a very large energy scale within a {\sl unification
approach}, which can be implemented in many extensions of the
SM. Alternatively, neutrino masses could arise from garden-variety
{\sl weak-scale physics} specified by a scale $\vev{\sigma} =
\O(m_Z)$ where $\vev{\sigma}$ denotes a \21 singlet vacuum expectation
value which owes its smallness to the symmetry enhancement which would
result if $\vev{\sigma}$ and $m_\nu \to 0$.

One should realize however that, the physics of neutrinos can be
rather different in various gauge theories of neutrino mass, and that
there is hardly any predictive power on masses and mixings, which
should not come as a surprise, since the problem of mass in general is
probably one of the deepest mysteries in present-day physics.

\subsection{Unification or Seesaw Neutrino Masses}
\vskip .1cm

The observed violation of parity in the weak interaction may be a
reflection of the spontaneous breaking of B-L symmetry in the context
of left-right symmetric extensions such as the \lr \cite{LR}, \422
\cite{PS} or \10 gauge groups \cite{GRS}. In this case the masses of
the light neutrinos are obtained by diagonalizing the following mass
matrix in the basis $\nu,\nu^c$
\begin{equation}
\left[\matrix{
 M_L & D \cr
 D^T & M_R }\right] 
\label{SS} 
\end{equation} 
where $D$ is the standard \21 breaking Dirac mass term and $M_R =
M_R^T$ is the isosinglet Majorana mass that may arise from a 126
vacuum expectation value in \10. The magnitude of the $M_L
\nu\nu$ term \cite{2227} is also suppressed by the left-right breaking
scale, $M_L \propto 1/M_R$ \cite{LR}.  In the seesaw approximation,
one finds
\beq 
M_{\nu \: eff} = M_L - D M_R^{-1} D^T
\:.
\label{SEESAW} 
\eeq 
As a result one is able to explain naturally the relative smallness of
\neu masses since $m_\nu \propto 1/M_R$.  Although $M_R$ is expected
to be large, its magnitude heavily depends on the model and it may
have different possible structures in flavour space (so-called
textures) \cite{Lola:1998xp}.  In general one can not predict the
corresponding light neutrino masses and mixings. In fact this freedom
has been exploited in model building in order to account for an almost
degenerate seesaw-induced neutrino mass spectrum \cite{DEG}.

One virtue of the unification approach is that it may allow one to
gain a deeper insight into the flavour problem. There have been
interesting attempts at formulating supersymmetric unified schemes
with flavour symmetries and texture zeros in the Yukawa couplings.  In
this context a challenge is to obtain the large lepton mixing now
indicated by the atmospheric neutrino data.

\subsection{Weak-Scale Neutrino Masses}
\vskip .1cm

Neutrinos may acquire mass from extra particles with masses $\O(m_Z)$
an therefore accessible to present experiments. There is a variety of
such mechanisms, in which neutrinos acquire mass either at the tree
level or radiatively.  Let us look at some examples, starting with the
tree level case.

\subsubsection{Tree-level Neutrino Masses}
\vskip .1cm

Consider the following extension of the lepton sector of the \21
theory: let us add a set of $two$ 2-component isosinglet neutral
fermions, denoted ${\nu^c}_i$ and $S_i$, $i=e,~\mu$ or $\tau$ in each
generation. In this case one can consider the mass matrix (in the
basis $\nu, \nu^c, S$) \cite{CON}
\begin{equation}
\left[\matrix{
  0 & D & 0 \cr
  D^T & 0 & M \cr
  0 & M^T & \mu }\right] 
\label{MATmu} 
\end{equation} 
The Majorana masses for the neutrinos are determined from
\beq
M_L = D M^{-1} \mu {M^T}^{-1} D^T
\label{33}
\eeq 
In the limit $\mu \to 0$ the exact lepton number symmetry is recovered
and will keep neutrinos strictly massless to all orders in
perturbation theory, as in the SM. The corresponding texture of the
mass matrix has been suggested in various theoretical models
\cite{WYLER}, such as superstring inspired models~\cite{SST}. In the
latter the zeros arise due to the lack of Higgs fields to provide the
usual Majorana mass terms.
The smallness of neutrino mass then follows from the smallness of
$\mu$. The scale characterizing $M$, unlike $M_R$ in the seesaw
scheme, can be low. As a result, in contrast to the heavy neutral
leptons of the seesaw scheme, those of the present model can be light
enough as to be produced at high energy colliders such as LEP
\cite{CERN} or at a future Linear Collider. The smallness of $\mu$ is
in turn natural, in t'Hooft's sense, as the symmetry increases when
$\mu \to 0$, i.e.  total lepton number is restored.  This scheme is a
good alternative to the smallness of neutrino mass, as it bypasses
the need for a large mass scale, present in the seesaw unification
approach. One can show that, since the matrices $D$ and $M$ are not
simultaneously diagonal, the leptonic charged current exhibits a
non-trivial structure that cannot be rotated away, even if we set $\mu
\equiv 0$.  The phenomenological implication of this, otherwise
innocuous twist on the SM, is that there is neutrino mixing despite
the fact that light neutrinos are strictly massless.  It follows that
flavour and CP are violated in the leptonic currents, despite the
masslessness of neutrinos. The loop-induced lepton flavour and CP
non-conservation effects, such as $\mu \to e + \gamma$~\cite{BER,3E},
or CP asymmetries in lepton-flavour-violating processes such as $Z \to
e \bar{\tau}$ or $Z \to \tau \bar{e}$~\cite{CP} are precisely
calculable. The resulting rates may be of experimental interest
\cite{ETAU,TTTAU,cernlfv}, since they are not constrained by the
bounds on neutrino mass, only by those on universality, which are
relatively poor.  In short, this is a conceptually simple and
phenomenologically rich scheme.

Another remarkable implication of this model is a new type of resonant
neutrino conversion mechanism \cite{massless0}, which was the first
resonant mechanism to be proposed after the MSW effect \cite{MSW}, in
an unsuccessful attempt to bypass the need for neutrino mass in the
resolution of the solar neutrino problem. According to the mechanism,
massless neutrinos and anti-neutrinos may undergo resonant flavour
conversion, under certain conditions. Though these do not occur in the
Sun, they can be realized in the chemical environment of supernovae
\cite{massless}. Recently it has been pointed out how they may provide
an elegant approach for explaining the observed velocity of pulsars
\cite{Grasso:1998tt}.
 
\subsubsection{Radiative Neutrino Masses}
\vskip .1cm

\begin{figure}[t]
\centerline{\protect\hbox{\psfig{file=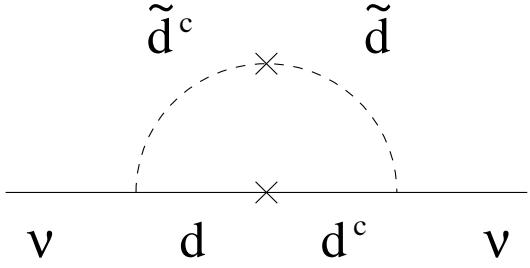,height=4cm,width=7cm}}}
\vglue -0.6cm
\caption{Mechanism for One-loop-induced Neutrino Mass. }
\label{mnrad}
\end{figure}

The prototype one-loop scheme is the one proposed by Zee
\cite{zee}. Supersymmetry with explicitly broken R-parity also
provides alternative one-loop mechanisms to generate neutrino mass
arising from scalar quark or scalar lepton exchanges, as shown in
\fig{mnrad}.

\begin{figure}
\centerline{\protect\hbox{\psfig{file=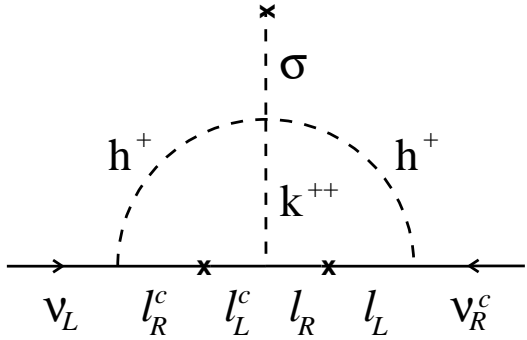,height=4.5cm,width=7cm}}}
\caption{Mechanism for Two-loop-induced Neutrino Mass }
\label{2loop}
\end{figure}

An interesting two-loop scheme to induce neutrino masses was suggested
by Babu \cite{Babu88}, based on the diagram shown in \fig{2loop}.
Note that I have used here a slight variant of the original model
which incorporates the idea of spontaneous~\cite{ewbaryo},  rather
than explicit lepton number violation.  

Finally, note also that one can combine these mechanisms as building
blocks in order to provide schemes for massive neutrinos. In
particular those in which there are not only the three active
neutrinos but also one or more light sterile neutrinos, such as those
in ref.~\cite{ptv92,pv93}.  In fact this brings in 
novel Feynman graph topologies.

\subsection{Supersymmetry: R-parity Violation as the Origin of Neutrino Mass}
\vskip .1cm

This is an interesting mechanism of neutrino mass generation which
combines seesaw and radiative mechanisms~\cite{epsrad}. It invokes
supersymmetry with broken R-parity, as the origin of neutrino mass and
mixings. The simplest way to illustrate the idea is to use the
bilinear breaking of R--parity~\cite{epsrad,RPothers} in a unified
minimal supergravity scheme with universal soft breaking parameters
(MSUGRA). Contrary to a popular misconception, the bilinear violation
of R--parity implied by the $\epsilon_3$ term in the superpotential is
physical, and can not be rotated away~\cite{BRpVtalks}. It leads also
by a minimization condition, to a non-zero sneutrino vev, $v_3$.  It
is well-known that in such models of broken R--parity the tau neutrino
$\nu_{\tau}$ acquires a mass, due to the mixing between neutrinos and
neutralinos~\cite{rossarca}. It comes from the matrix
\begin{equation}
\left[\matrix{
M_1 & 0  & -\half g'v_d & \half g'v_u & -\half g'v_3 \cr
0   & M_2 & \half g v_d & -\half g v_u & \half g v_3 \cr
-\half g'v_d & \half g v_d & 0 & -\mu & 0 \cr
\half g'v_u & -\half g v_u & -\mu & 0 & \epsilon_3 \cr
-\half g'v_3 & \half g v_3 & 0 & \epsilon_3 & 0 
}\right]
\label{eq:NeutMassMat}
\end{equation}
where the first two rows are gauginos, the next two Higgsinos, and the
last one denotes the tau neutrino. The $v_u$ and $v_d$ are the
standard vevs, $g's$ are gauge couplings and $M_{1,2}$ are the gaugino
mass parameters. Since the $\epsilon_3$ and the $v_3$ are related, the
simplest (one-generation) version of this model contains only one
extra free parameter in addition to those of the MSUGRA model.  The
universal soft supersymmetry-breaking parameters at the unification
scale $m_X$ are evolved via renormalization group equations down to
the weak scale $\O(m_Z)$.  This induces an effective non-universality
of the soft terms {\sl at the weak scale} which in turn implies a
non-zero sneutrino vev $v'_3$ given as
\begin{equation}
v'_3 \approx \frac{\epsilon_3 \mu} {{m_Z}^4}
\left(v'_d \Delta M^2 + \mu'v_u \Delta B \right)
\label{App_v3p}
\end{equation}
where the primed quantities refer to a basis in which we eliminate the
$\epsilon_3$ term from the superpotential (but reintroduce it, of
course, in other sectors of the theory).  

The scalar soft masses and bilinear mass parameters obey $\Delta
M^2=0$ and $\Delta B=0$ at $m_X$. However at the weak scale they are
calculable from radiative corrections as
\begin{eqnarray}
\Delta M^2 & \approx & {{3h_b^2} \over{8\pi^2}} m_{Z}^2
\ln{{M_{GUT}}\over{m_Z}}
\end{eqnarray}
Note that \eq{App_v3p} implies that the R--parity-violating effects
induced by $v'_3$ are {\sl calculable} in terms of the primordial
R--parity-violating parameter $\epsilon_3$. It is clear that the
universality of the soft terms plays a crucial r\^ole in the
calculability of the $v'_3$ and hence of the resulting neutrino mass
\cite{epsrad}. Thus \eq{eq:NeutMassMat} represents a new kind of
see-saw scheme in which the $M_R$ of \eq{SS} is the neutralino mass,
while the r\^ole of the Dirac entry $D$ is played by the $v'_3$, which
is induced radiatively as the parameters evolve from $m_X$ to the weak
scale. Thus we have a {\sl hybrid} see-saw mechanism, with naturally
suppressed Majorana $\nu_{\tau}$ mass induced by the mixing between
the weak eigenstate tau neutrino and the {\sl zino}.

\begin{figure}[t]
\centerline{\protect\hbox{\psfig{file=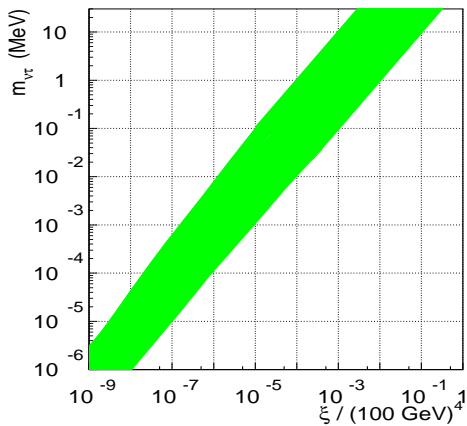,height=6cm,width=7cm}}}
\caption{Tau neutrino mass from Broken R--parity vs 
$\xi \equiv(\epsilon_3v_d+\mu v_3)^2$ from ref.~\protect\cite{epsrad}.
}
\label{mnt_xi_ev}
\end{figure}

In order to estimate the expected \nt mass let me first determine the
tau neutrino mass in the most general supersymmetric model with
bilinear breaking of R-parity, {\sl without imposing universality of
the soft SUSY breaking terms}.  The \nt mass depends quadratically on
an effective parameter $\xi$ defined as $\xi \equiv (\epsilon_3 v_d +
\mu v_3)^2 \propto {v'_3}^2$ characterizing the violation of R--parity.  
The expected \mnt values are illustrated in \fig{mnt_xi_ev}. The band
shown in the figure is obtained through a scan over the parameter
space requiring that the supersymmetric particles are not too light.
This should be compared with the cosmologically allowed values of the
tau neutrino mass $\sum m_\nu \lsim 92 \Omega h^2$ eV (see below).
Note that this only holds if neutrinos are stable.  In the present
model the \nt is expected to decay into 3 neutrinos, via the neutral
current \cite{2227,774}, or by slepton exchanges. This decay will
reduce the relic \nt abundance to the required level, as long as \nt
is heavier than about 200 KeV or so. Since on the other hand
primordial Big-Bang nucleosynthesis implies that \nt is lighter than
about an MeV or so~\cite{bbnutaustable} there is a forbidden gap in
this model if the majoron is not introduced. In the full version of
the model the presence of the majoron allows all neutrino masses to be
viable cosmologically.

Back to the simplest model with explicit bilinear breaking of
R--parity, let me note that in this model the \nt mass can be very large. A
way to obtain a model with a small and calculable \nt mass, as
indicated by the simplest interpretation of the atmospheric neutrino
anomaly in terms of \nm to \nt oscillations is to assume a SUGRA
scheme with universality of the soft supersymmetry breaking terms at
$m_X$. In this case the \nt mass is theoretically {\sl predicted} in
terms of $h_b$ and can be small in this case due to a natural
cancellation between the two terms in the parameter $\xi$, which
follows from the assumed universality of the soft terms at $m_X$. One
can verify that \mnt may easily lie in the ten electron-volt
range. Lower masses require about two orders of magnitude in addition
to that which is dictated by the RGE evolution, which is certainly
not unreasonable.  Moreover the solution of the atmospheric neutrino
anomaly may involve some exotic mechanism, such as the FC interactions
\cite{Gonzalez-Garcia:1998hj}.

As a last remark I note that \ne and \nm remain massless in this
approximation. They get masses either from scalar loop contributions
in \fig{mnrad} or by mixing with singlets in models with spontaneous
breaking of R-parity~\cite{Romao92}.  A detailed study is now underway
of the loop contributions to \nm and \ne is underway in Valencia.  It
is important to notice that even when \mnt is small, many of the
corresponding R-parity violating effects can be sizeable. An obvious
example is the fact that the lightest neutralino decay will typically
decay inside the detector, unlike standard R-parity-conserving
supersymmetry. This leads to a vastly unexplored plethora of
phenomenological possibilities in supersymmetric physics
\cite{desert}.


In conclusion one can see that, of the various attractive schemes for
giving neutrinos a mass, only the seesaw scheme requires a large mass
scale. It gives a grand connection between the very light (the
neutrinos) and the very heavy (some unknown particles). At this stage
is is premature to bet on any mechanism and from this point of view
neutrinos open the door to a potentially rich phenomenology, since the
extra particles required have masses at scales that could be
accessible to present experiments.  In the simplest versions of these
models the neutrino mass arises from the explicit violation of lepton
number. Their phenomenological potential gets even richer if one
generalizes the models so as to implement a spontaneous violation
scheme. This brings me to the next section.

\subsection{Majorons at the Weak-scale}
\vskip .1cm

The generation of neutrino masses will be accompanied by the existence
of a physical Goldstone boson that we generically call majoron in any
model where lepton number (or B-L) is an ungauged symmetry which is
arranged to break spontaneously.
Except for the left-right symmetric unification approach, in which B-L
is a gauge symmetry, in all of the above schemes one can implement the
spontaneous violation of lepton number. 
One can also introduce it in a seesaw framework, both with \21
\cite{CMP} as well as left-right symmetry \cite{Akhmedov:1995wd}.
While in the \21 case it is rather simple, in the case of
left-right-symmetric models, one needs to implement a spontaneously
broken global U(1) symmetry similar to lepton number. One interesting
aspect that emerges in the latter case is that it allows also the
left-right scale to be relatively low~\cite{Akhmedov:1995wd}.  Here I
do not consider the seesaw-type majorons, for a discussion see
ref.~\cite{fae}. I will mainly concentrate on weak-scale physics. In
all models I consider the lepton-number breaks at a scale given by a
vacuum expectation value $\vev{\sigma} \sim m_{weak}$.
In all of these models, the weak scale arises as the most natural one
and, as already mentioned, the neutrino masses when $\vev{\sigma} \to
0$ i.e. when the lepton-breaking scale vanish \cite{JoshipuraValle92}.

In any phenomenologically acceptable model one must arrange for the
majoron to be mainly an \21 singlet, ensuring that it does not affect
the invisible Z decay width, well-measured at LEP. In models where the
majoron has L=2 the neutrino mass is proportional to an insertion of
$\vev{\sigma}$, as indicated in \fig{2loop}.  In the supersymmetric
model with broken R-parity the majoron is mainly a singlet sneutrino,
which has lepton number L=1, so that $m_\nu \propto \vev{\sigma}^2$,
where $\vev{\sigma} \equiv \vev{\widetilde{\nu^c}}$, with
$\widetilde{\nu^c}$ denoting the singlet sneutrino. The presence of
the square, just as in the parameter $\xi$ ~in \fig{mnt_xi_ev},
reflects the fact that the neutrino gets a Majorana mass which has
lepton number L=2.  The sneutrino gets a vev at the effective
supersymmetry breaking scale $ m_{susy} = m_{weak}$.

The weak-scale majorons may have other phenomenological
implications. One is the possibility of invisibly decaying Higgs
bosons~\cite{JoshipuraValle92} which I have no time to discuss here
(see, for instance \cite{desert}).

Finally note that if the majoron acquires a KeV mass (natural in
weak-scale models) from gravitational effects at the Planck scale
\cite{ellis} it may play obey the main requirements to play a 
r\^ole in cosmology as dark matter~\cite{KEV}.
In what follows I will just focus on two examples of how the
underlying physics of weak-scale majoron models can affect neutrino
cosmology in an important way.

\subsubsection{Heavy  Neutrinos and the Universe Mass}
\vskip .1cm

Neutrinos of mass less than \O(100 KeV) or so, are cosmologically
stable if they have only SM interactions. Their contribution to the
present density of the universe implies \cite{KT}
\beq 
\label{RHO1}
\sum m_{\nu_i} \lsim 92 \: \Omega_{\nu} h^2 \: eV\:, 
\eeq 
where the sum is over all isodoublet neutrino species with mass less
than \O(1 MeV). The parameter $\Omega_{\nu} h^2 \leq 1$, where $h^2$
measures the uncertainty in the present value of the Hubble parameter,
$0.4 \lsim h \lsim 1$, while $\Omega_{\nu} = \rho_{\nu}/\rho_c$,
measures the fraction of the critical density $\rho_c$ in neutrinos.
For the $\nu_{\mu}$ and $\nu_{\tau}$ this bound is much more stringent
than the laboratory limits.

In weak-scale majoron models the generation of neutrino mass is
accompanied by the existence of a physical majoron, which leads to
potentially fast majoron-emitting decay channels such as \cite{fae,V}
\beq
\label{NUJ}
\nu^\prime  \to \nu + J \:\: .
\eeq
as well as new annihilations to majorons,
\beq
\label{nunuJJ}
\nu^\prime  + \nu^\prime  \to J + J \:\: .
\eeq
These could eliminate relic neutrinos and therefore allow neutrinos of
higher mass, as long as the rates are large enough to allow for an
adequate red-shift of the heavy neutrino decay and/or annihilation
products. While the annihilation involves a diagonal majoron-neutrino
coupling $g$, the decays proceed only via the non-diagonal part of the
coupling, in the physical mass basis.  A careful diagonalization of
both mass matrix and coupling matrix is essential in order to avoid
wild over-estimates of the heavy neutrino decay rates, such as that in
ref.~\cite{CMP}.
The point is that, once the neutrino mass matrix is diagonalized,
there is a danger of simultaneously diagonalizing the majoron
couplings to neutrinos. That would be analogous to the GIM mechanism
present in the SM for the couplings of the Higgs to fermions. Models
that avoid this GIM mechanism in the majoron-neutrino couplings have
been proposed, e.g. in ref.~\cite{V}. Many of them are weak-scale
majoron models \cite{CON,Romao92,JoshipuraValle92}.  A general method
to determine the majoron couplings to neutrinos and hence the neutrino
decay rates in any majoron model was first given in ref. \cite{774}.
For an estimate in the model with spontaneously broken R-parity
\cite{MASIpot3} see ref. \cite{Romao92}.

One can summarize that since neutrinos can be short-lived their masses
can only be really constrained by laboratory experiments based on
direct search. The cosmological and other bounds are important but
require additional theoretical elements in their interpretation.

\subsubsection{Heavy Neutrinos and  Cosmological Nucleosynthesis} 
\vskip .1cm

\begin{figure}
\centerline{\protect\hbox{\psfig{file=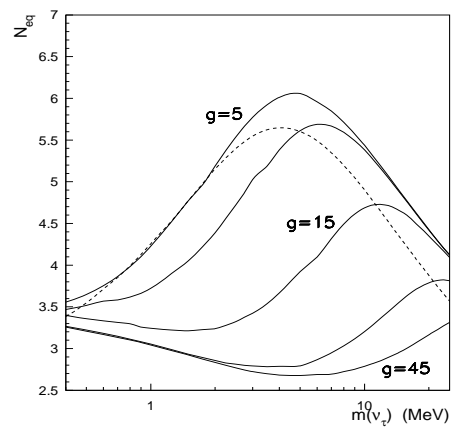,height=5.6cm,width=6cm}}}
\caption{Effective number of massless SM neutrinos equivalent to the 
heavy \nt, as given in ref. \protect\cite{DPRV}. Non-zero $g$ values
(in units of $10^{-5}$) can lower $N_\nu$ with respect to the SM case
$g=0$ (dashed line) due to the effect of annihilations. }
\label{bbneq}
\end{figure}

The number of light neutrino species is restricted by cosmological Big
Bang Nucleosynthesis (BBN). Due to its large mass, an MeV stable
(lifetime longer than $\sim 100$ sec) tau neutrino would be equivalent
to several SM massless neutrino species and would therefore
substantially increase the abundance of primordially produced
elements, e.g. $^{4}He$ and deuterium
\cite{dmeas,cris.ncris,sarkar,Fiorentini:1998fv}.  This can be converted 
into restrictions on the \nt mass. If the bound on the effective
number of massless neutrino species is taken as $N_\nu < 3.4-3.6$, one
can rule out $\nu_\tau$ masses above 0.5 MeV~\cite{bbnutaustable}.  If
we take $N_\nu < 4.5$ \cite{sarkar} the \mnt limit loosens
accordingly, as seen from \fig{bbneq}, and allows a \nt of about an
MeV or so.

\begin{figure}
\centerline{
\psfig{file=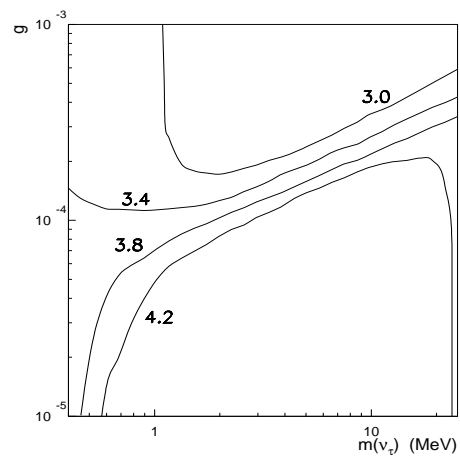,height=6cm,width=6cm}}
\caption{The BBN-allowed regions for each $N_\nu^{max}$ lie above the
respective curve \protect\cite{DPRV} }
\label{neffmg}
\end{figure}

In the presence of \nt annihilations the BBN \mnt bound is
substantially weakened or eliminated \cite{DPRV}. In \fig{bbneq} we
also give the expected $N_\nu$ value for different values of the
coupling $g$ between $\nu_\tau$'s and $J$'s, expressed in units of
$10^{-5}$. 
Comparing with the SM $g=0$ case one sees that for a fixed
$N_\nu^{max}$, a wide range of tau neutrino masses is allowed for
large enough values of $g$. No \nt masses below the LEP limit can be
ruled out, as long as $g$ exceeds a few times $10^{-4}$.
One can also see from the figure that {\sl $N_\nu$ can also be lowered
below the canonical SM value $N_\nu = 3$} due to the effect of the
heavy \nt annihilations to majorons.
These results may be re-expressed in the $m_{\nu_\tau}-g$ plane, as
shown in figure \ref{neffmg}.  
We note that the required values of $g(m_{\nu_\tau})$ fit well with
the theoretical expectations of many weak-scale majoron models.

As we have seen \nt annihilations to majorons may weaken or even
eliminate the BBN constraint on the tau neutrino mass. Similarly, in
some weak-scale majoron models decays in \eq{NUJ} may lead to short
enough \nt lifetimes that they may also play an important r\^ole in
BBN, again with the possibility of substantially weakening or
eliminating the BBN constraint on the tau neutrino mass
\cite{bbnunstable}.

\section{Indications for New Physics}
\vskip .1cm

The most solid hints in favour of new physics in the neutrino sector
come from underground experiments on solar and atmospheric
neutrinos. The published
data~\cite{Fukuda:1999ua,Fukuda:1999rq,Fukuda:1998mi} data correspond
to a 504--day solar neutrino data sample~\cite{sk504s} and 535--day
atmospheric neutrino data sample~\cite{sk535a}, respectively. These
were the data first presented at the past Neutrino 98 conference in
Japan.  Here we include also some results from the more recent
708--day data sample, see ref. \cite{Smy:1999tt}
and~\cite{Inoue99venice}.

\subsection{Solar Neutrinos}
\vskip .1cm
 
The data collected by the Kamiokande, and the radiochemical Homestake,
Gallex and Sage experiments have no Standard Model explanation. The
event rates are summarized as: $2.56 \pm 0.23$ SNU (chlorine), $72.2
\pm 5.6$ SNU (Gallex and Sage gallium experiments sensitive to the
$pp$ neutrinos), and $(2.44 \pm 0.10) \times 10^6 {\rm cm^{-2} s^{-1}
}$ ($^8$B flux from Super-Kamiokande)~\cite{solarexp}.  
\begin{figure}[t]
\centerline{\protect\hbox{
\psfig{file=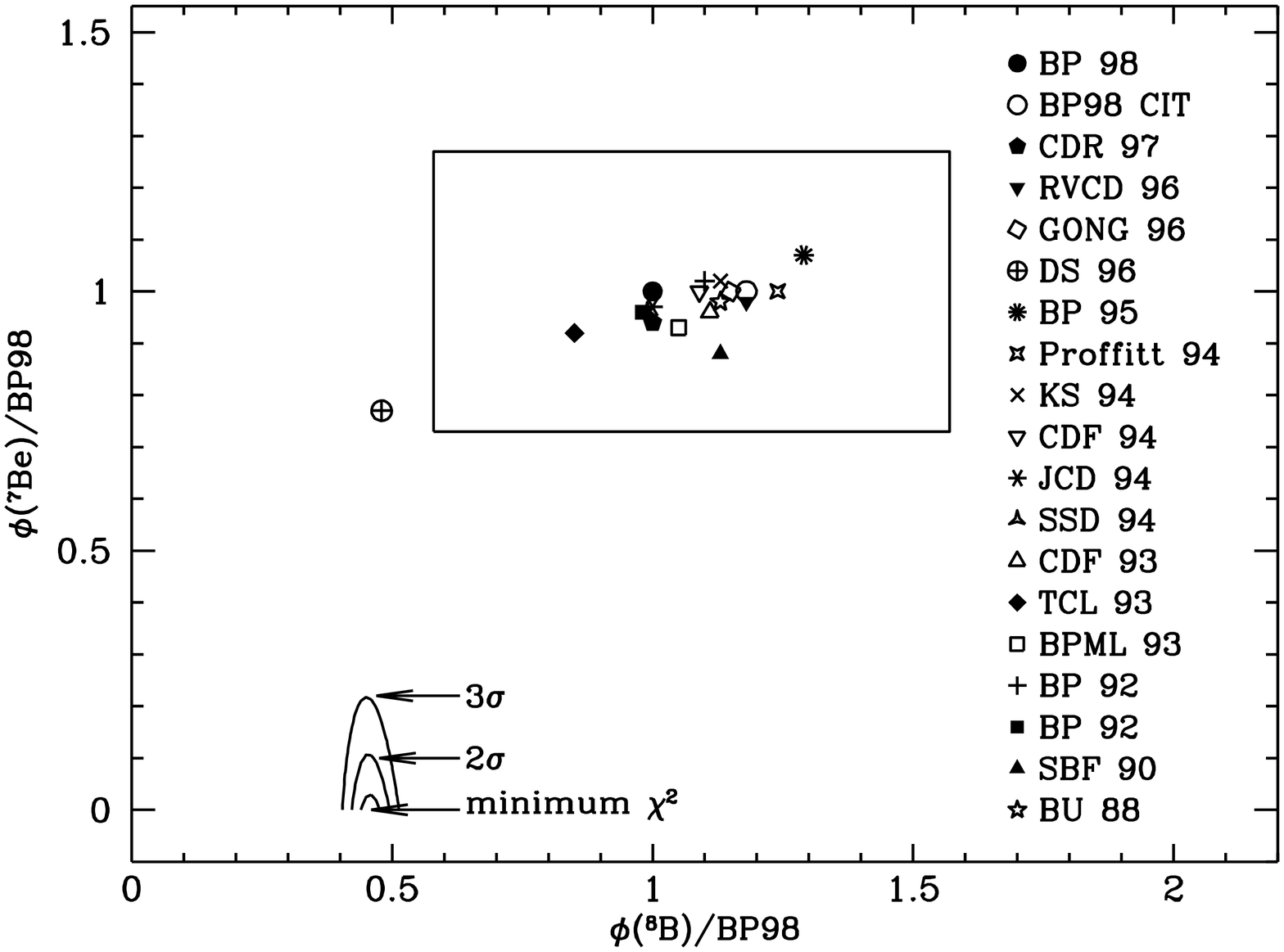,height=6cm,width=6cm}}}
\caption{SSM predictions, from ref.~\protect\cite{Bahcall98}}
\label{78}
\end{figure}
In \fig{78} one can see the predictions of various standard solar
models in the plane defined by the $^7$Be and $^8$B neutrino fluxes,
normalized to the predictions of the BP98 solar model~\cite{BP98}.
Abbreviations such as BP95, identify different solar models, as given
in ref.~\cite{models}.  The rectangular error box gives the $3\sigma$
error range of the BP98 fluxes. The values of these fluxes indicated
by present data on neutrino event rates are also shown by the contours
in the figure. The best-fit $^7$Be neutrino flux is negative!
Possible non-standard astrophysical solutions are strongly constrained
by helioseismology studies \cite{Bahcall98,helio97}. Within the
standard solar model approach, the theoretical predictions clearly lie
well away from the $3\sigma$ contour, strongly suggesting the need for
new particle physics in order to account for the data~\cite{CF}.

The most likely possibility is to assume the existence of neutrino
conversions, such as could be induced by very small neutrino masses.
Possibilities include the MSW effect~\cite{MSW}, vacuum neutrino
oscillations \cite{Glashow:1987jj,bimax98,Mohapatra:1986ks}, the
Resonant~\cite{RSFP,akhmedov97} Spin-Flavour Precession mechanism
\cite{SFP} and, possibly, flavour changing neutrino
interactions \cite{Krastev:1997cp}.

The recent 708--day data sample~\cite{Smy:1999tt} presents no major
surprises, except that the recoil energy spectrum produced by solar
neutrino interactions shows more events in the highest bins.  Barring
the possibly of poorly understood energy resolution effects, Bahcall
and Krastev~\cite{bkhep} have noted that if the flux for neutrinos
coming from the ${\rm ^3He} ~+~ p \to {\rm ^4He} ~+~e^+ ~+~\nu_e $,
the so-called $hep$ reaction, is well above the (uncertain) SSM
predictions, then this could significantly influence the electron
energy spectrum produced by solar neutrino interactions in the high
recoil region, with hardly any effect at lower energies.
\begin{figure}
\centerline{\protect\hbox{
\psfig{file=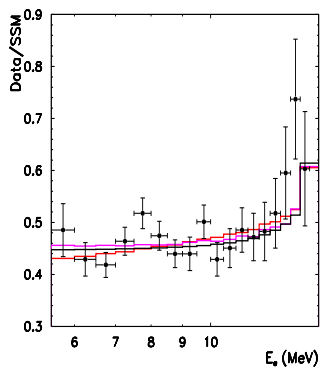,height=7cm,width=6cm}}}
\vglue -.5cm
\caption{Expected normalized recoil electron
energy spectra versus 708--day SK data,
from~\protect\cite{valencia:1999}.  }
\label{spec708}
\end{figure}
Fig. \ref{spec708} shows the expected normalized recoil electron
energy spectrum compared with the most recent experimental
data~\protect\cite{Smy:1999tt}. The solid line represents the
prediction for the best--fit SMA solution with free $^8B$ and $hep$
normalizations (0.69 and 12 respectively), while the dotted line gives
the corresponding prediction for the best--fit LMA solution (1.15 and
34 respectively). Finally, the dashed line represents the prediction
for the best no-oscillation scheme with free $^8B$ and $hep$
normalizations (0.44 and 14, respectively). Clearly the spectra with
enhanced $hep$ neutrinos provide better fits to the data. However
Fiorentini et al~\cite{Fiorentini:1998xr} have argued that the
required $hep$ amount is too large to accept on theoretical
grounds. We look forward to the improvement of the situation in the
next round of data.  The increasing r\^ole played rate-independent
observables such as the spectrum, as well as seasonal and day-night
asymmetries, marks a turning point in solar neutrino research, which
will eventually select the mechanism responsible for the explanation
of the solar neutrino problem.

\begin{figure}
\centerline{
\protect\hbox{\psfig{file=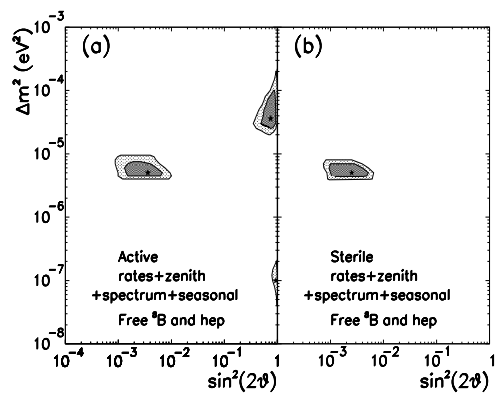,width=7cm,height=6cm}}}
\vglue -.5cm
\caption{Solar neutrino parameters for 2-flavour MSW neutrino 
conversions with 708--day SK data sample and enhanced $hep$ flux
~\protect\cite{valencia:1999}}
\label{msw}
\end{figure}

The required solar neutrino parameters are determined through a
$\chi^2$ fit of the experimental data.  In~\fig{msw} we show the
allowed regions in $\Delta m^2$ and $\sin^2\theta$ from the
measurements of the total event rates at the Chlorine, Gallium and
Super--Kamiokande (708-day data sample) combined with the zenith angle
distribution observed in Super--Kamiokande, the recoil energy spectrum
and the seasonal dependence of the event rates, for active-active
oscillations {\bf(a)} and active-sterile oscillations {\bf(b)} .  The
darker (lighter) areas indicate 90\% (99~\%)CL regions.  The best--fit
points in each region are indicated by a star.  The analysis uses free
$^8B$ and $hep$ normalizations~\cite{valencia:1999}
One notices from the analysis that rate-independent observables, such
as the electron recoil energy spectrum and the day-night asymmetry
(zenith angle distribution), are playing an increasing r\^ole in
ruling out large regions of parameters~\cite{valencia:1999}.  Another
example of an observable which has been neglected in most analyses of
the MSW effect and which could be sizeable for the large mixing angle
(LMA) region is the seasonal dependence in the solar neutrino flux
which would result from the regeneration effect at the Earth and which
has been discussed in ref.~\cite{deHolanda:1999ty}. This should play a
more significant r\^ole in future investigations.

A theoretical issue which has raised some interest recently is the
study of the possible effect of random fluctuations in the solar
matter density \cite{BalantekinLoreti,noise,noise2}. The possible
existence of noise fluctuations at a few percent level is not excluded
by present helioseismology studies. 
\begin{figure}
\centerline{\protect\hbox{\psfig{file=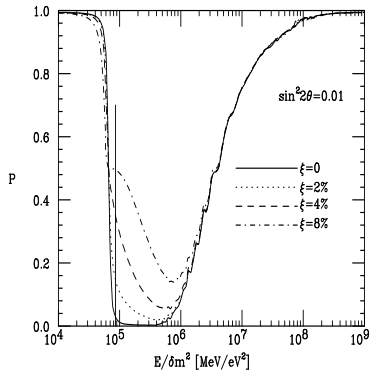,width=6cm,height=6cm,angle=90}}}
\vglue -.5cm
\caption{ Solar neutrino survival probability in a noisy Sun, 
from ref.~\protect\cite{noise}}
\label{Pnoise}
\end{figure}
In \fig{Pnoise} we show averaged solar neutrino survival probability
as a function of $E/\Delta m^2$, for $\sin^2 2\theta = 0.01$. This
figure was obtained via a numerical integration of the MSW evolution
equation in the presence of noise, using the density profile in the
Sun from BP95 in ref.~\cite{models}, and assuming that the correlation
length $L_0$ (which corresponds to the scale of the fluctuation) is
$L_0 = 0.1 \lambda_m$, where $\lambda_m$ is the neutrino oscillation
length in matter. An important assumption in the analysis is that $
l_{free} \ll L_0 \ll \lambda_m$, where $l_{free} \sim 10 $ cm is the
mean free path of the electrons in the solar medium. The fluctuations
may strongly affect the $^7$Be neutrino component of the solar
neutrino spectrum so that the Borexino experiment should provide an
ideal test, if sufficiently small errors can be achieved. The
potential of Borexino in probing the level of solar matter density
fluctuations provides an additional motivation for the experiment
\cite{borexino}. 
\begin{figure}
\centerline{\protect\hbox{\psfig{file=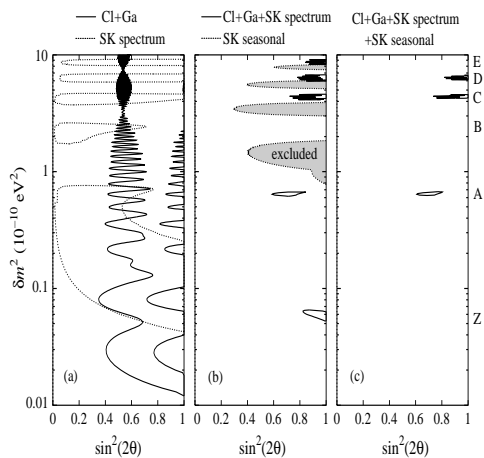,width=6.5cm,height=6cm}}}
\caption{Vacuum oscillation parameters, from ref.~\protect\cite{jso700}}
\label{vac99}
\end{figure}

The most popular alternative solution to the solar neutrino problem is
the {\sl vacuum oscillation solution} \cite{Glashow:1987jj} which
clearly requires large neutrino mixing and to adjust the
oscillation length so as to coincide roughly with the Earth-Sun
distance. This solution fits well with some theoretical models
\cite{Mohapatra:1986ks}.
Fig. \ref{vac99} shows the regions of just-so oscillation parameters
at the 95~\%~CL obtained in a recent fit of the data, including both
the rates, the recoil energy spectrum and seasonal effects, which are
expected in this scenario ~\cite{lisi} and could potentially help in
discriminating it from the MSW scenario.

\subsection{Atmospheric Neutrinos}
\vskip .1cm

There has been a long-standing discrepancy between the predicted and
measured \nm/\ne ratio \cite{atmexp} of the fluxes of atmospheric
neutrinos~\cite{atmreview}.  The anomaly was found both in water
Cerenkov experiments, Kamiokande, Super-Kamiokande and IMB
\cite{sk300}, as well as in the iron calorimeter Soudan2 experiment. 
Negative experiments, such as Frejus and Nusex have much larger
errors.

Although individual $\nu_\mu$ or $\nu_e$ fluxes are only known to
within $30\%$ accuracy, the $\nu_\mu$ $/\nu_e$ ratio is known to
$5\%$.  The most important feature of the atmospheric neutrino
535-day~ data~sample~\cite{sk535a} is that it exhibits a {\sl
zenith-angle-dependent} deficit of muon neutrinos which is
inconsistent with theoretical expectations. For recent analyses see
ref.~\cite{atm98,atmo98}. Experimental biases and uncertainties in the
prediction of neutrino fluxes and cross sections are unable to explain
the data.

\begin{figure}
\centerline{\protect\hbox{\epsfig{file=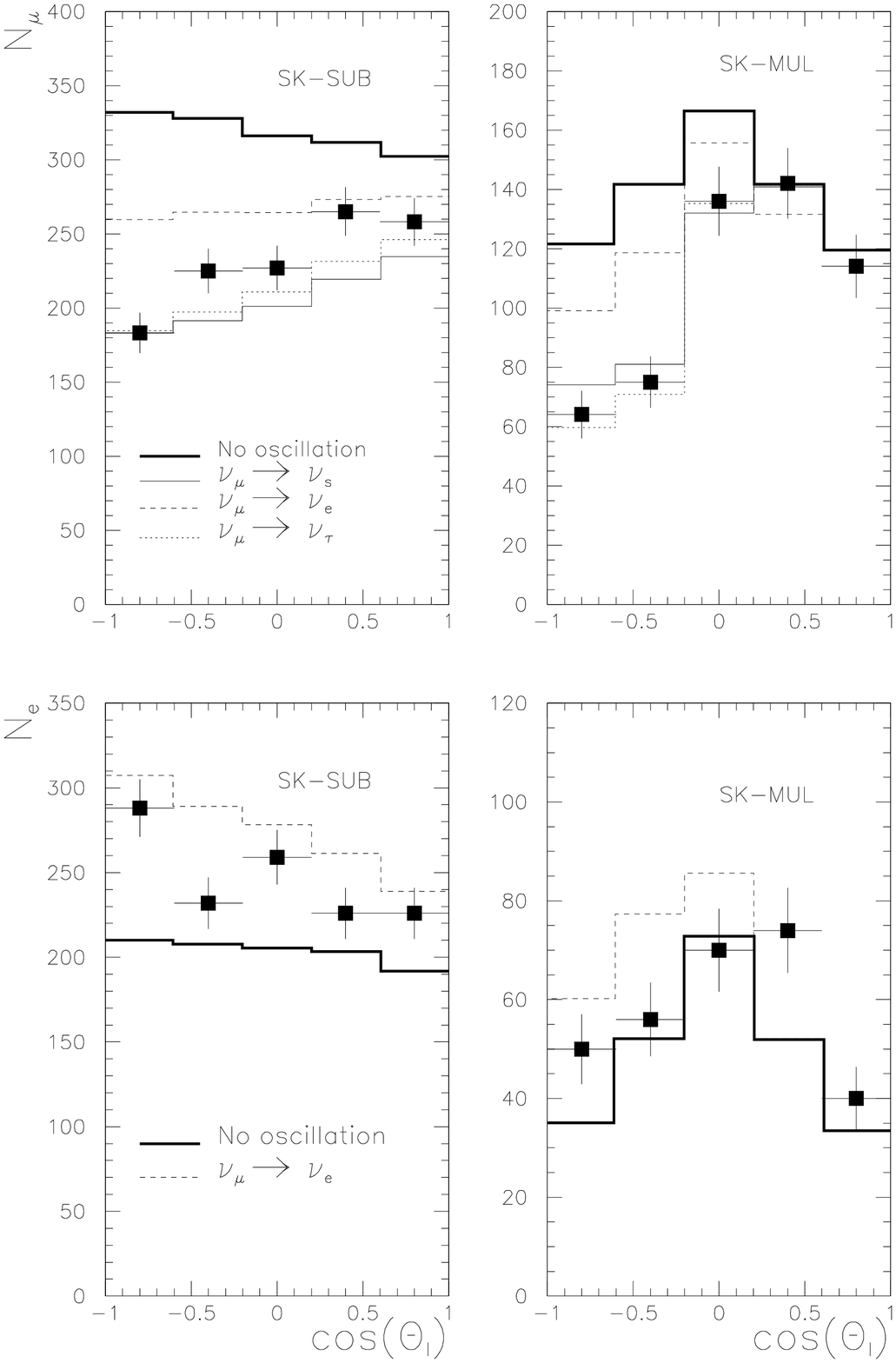,width=7cm,height=8cm}}}
\caption{Expected zenith angle distributions for SK
electron and muon-like sub-GeV and multi-GeV events in the SM
(no-oscillation) and for the best-fit points of the various
oscillation channels, from ref.~\protect\cite{atm98}. }
\label{ang_mu}  
\end{figure}

Fig. \ref{ang_mu} shows the measured zenith angle distribution of
electron-like and muon-like sub-GeV and multi-GeV events, as well as
the one predicted in the absence of oscillation. It also gives the
expected distribution in various neutrino oscillation schemes.
The thick-solid histogram is the theoretically expected distribution
in the absence of oscillation, while the predictions for the best-fit
points of the various oscillation channels is indicated as follows:
for $\nu_\mu \to \nu_s$ (solid line), $\nu_\mu \to \nu_e$ (dashed
line) and $\nu_\mu \to \nu_\tau$ (dotted line).  The error displayed
in the experimental points is only statistical.  The analysis used the
latest improved calculations of the atmospheric neutrino fluxes as a
function of zenith angle, including the muon polarization effect and
took into account a variable neutrino production point
\cite{flux}.  

\begin{figure}
\centerline{\protect\hbox{\epsfig{file=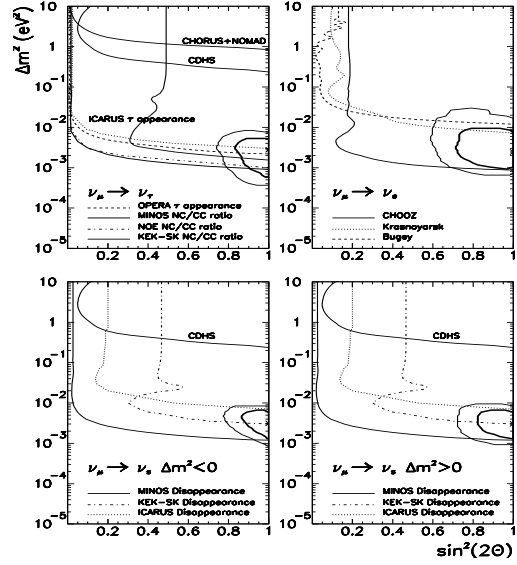,width=7.8cm,height=8cm}}}
\caption{Allowed atmospheric oscillation parameters for all
experiments including the 535-day SK  data, combined at
90 (thick solid line) and 99 \% CL (thin solid line) for all possible
oscillation channels, from ref.~\protect\cite{atm98}.  In each case
the best-fit point is denoted by a star and always corresponds to
maximal mixing, a feature which is well-reproduced by the theoretical
predictions of the models proposed in
ref.~\protect\cite{ptv92,pv93}. The sensitivity of the present
accelerator and reactor experiments as well as the expectations of
upcoming long-baseline experiments is also displayed.}
\label{mutausk4} 
\end{figure}

Clearly the data are not reproduced by the no-oscillation
hypothesis. The most popular way to account for this anomaly is in
terms of neutrino oscillations. In~\fig{mutausk4} I show the allowed
parameters obtained in a global fit of the sub-GeV and multi-GeV
(vertex-contained) atmospheric neutrino data~\cite{atm98} including
the 535 day SK  data, as well as all other experiments
combined at 90 (thick solid line) and 99 \% CL (thin solid line) for
each oscillation channel considered.
The two lower panels \fig{mutausk4} differ in the sign of the $\Delta
m^2$ which was assumed in the analysis of the matter effects in the
Earth for the $\nu_\mu \to \nu_s$ oscillations. Though $\nu_\mu
\to \nu_\tau$ oscillations give a slightly better fit than $\nu_\mu
\to \nu_s$ oscillations,  at present the atmospheric neutrino data
cannot distinguish between these channels.
It is well-known that the neutral-to-charged current ratios are
important observables in neutrino oscillation phenomenology, which are
especially sensitive to the existence of singlet neutrinos, light or
heavy \cite{2227}.
The atmospheric neutrinos produce isolated neutral pions
($\pi^0$-events) mainly in neutral current interactions.
One may therefore study the ratios of $\pi^0$-events and the events
induced mainly by the charged currents, as recently advocated in
ref.~\cite{vissani}. This minimizes uncertainties related to the
original atmospheric neutrino fluxes.
In fact the Super-Kamiokande collaboration has estimated the double
ratio of $\pi^0$ over e-like events in their sample~\cite{sk535a} and
found $R = 0.93 \pm 0.07 \pm 0.19$. This is consistent both with \nm to \nt or
\nm to \ns channels, with a slight preference for the former. The
situation should improve in the future.  We also display in
\fig{mutausk4} the sensitivity of present accelerator and reactor
experiments, as well as that expected at future long-baseline (LBL)
experiments.  The first point to note is that the Chooz reactor
\cite{Chooz} data excludes the region indicated for the $\nu_{\mu} \to
\nu_e$ channel when all experiments are combined at 90\% CL.

From the upper-left panel in \fig{mutausk4} one sees that the regions
of $\nu_\mu \to \nu_\tau$ oscillation parameters obtained from the
atmospheric neutrino data analysis cannot be fully tested by the LBL
experiments, as presently designed.  One might expect that, due to the
upward shift of the $\Delta m^2$ indicated by the fit for the sterile
case (due to the effects of matter in the Earth) it would be possible
to completely cover the corresponding region of oscillation
parameters. Although this is the case for the MINOS disappearance
test, in general most of the LBL experiments can not completely probe
the region of oscillation parameters indicated by the $\nu_\mu \to
\nu_s$ atmospheric neutrino analysis, irrespective of the
sign of $\Delta m^2$ assumed.  For a discussion of the various
potential tests that can be performed at the future LBL experiments in
order to unravel the presence of oscillations into sterile channels
see ref.~\cite{atmconcha}.

However appealing it may be, the neutrino oscillation interpretation
of the atmospheric neutrino anomaly is at the moment by no means
unique.  Indeed, the anomaly can be well accounted for in terms of
flavour changing neutrino interactions, with no need for neutrino mass
or mixing~\cite{Gonzalez-Garcia:1998hj}.  Investigations involving
upward through going muons by Superkamiokande \cite{Fukuda:1998ah} as
well as other experiments will play an important r\^ole in
discriminating between oscillations and alternative mechanisms to
explain the sub and multi-GeV atmospheric neutrino
data~\cite{fornengo99}.

\subsection{Other Hints}

\subsubsection{LSND}

\begin{figure}
\centerline{\protect\hbox{\epsfig{file=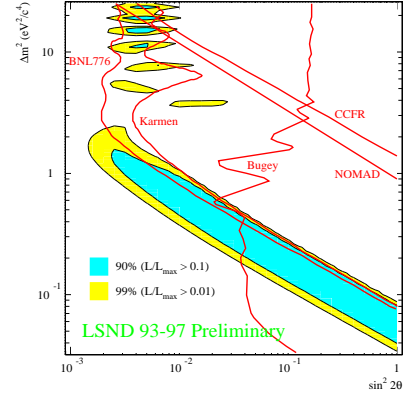,width=6cm,height=6cm}}}
\caption{Allowed LSND oscillation parameters versus competing
experiments~\protect\cite{louis} }
\label{darlsnd} 
\vglue -.5cm
\end{figure}

\begin{figure}
\centerline{\protect\hbox{\epsfig{file=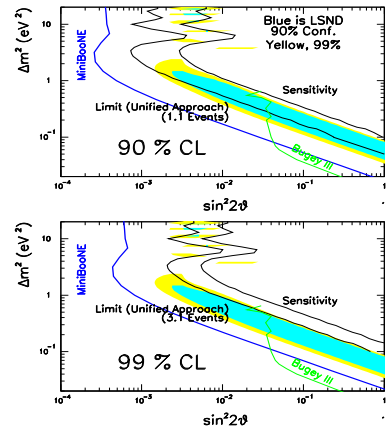,width=6cm,height=6cm}}}
\caption{Expected sensitivity of the proposed MiniBooNE
experiment~\protect\cite{louis} }
\label{miniboone} 
\end{figure}

The Los Alamos Meson Physics Facility  looked for
$\bar\nu_{\mu}\to \bar\nu_{e}$ oscillations using $\bar\nu_\mu$ from
$\mu^+$ decay at rest \cite{LSND}. The $\bar\nu_e$'s are detected via
the reaction $\bar\nu_e\,p \to e^{+}\,n$, correlated with a $\gamma$
from $np \to d \gamma$ ($2.2\,{\rm MeV}$).  The results indicate $\bar
\nu_\mu \to \bar \nu_e$ oscillations, with an oscillation probability 
of ($0.31^{+0.11}_{-0.10} \pm 0.05$)\%, leading to the oscillation
parameters shown in~\fig{darlsnd}. The shaded regions are the favoured
likelihood regions given in ref.~\cite{LSND}.  The curves show the
90~\% and 99~\% likelihood allowed ranges from LSND, and the limits
from BNL776, KARMEN1, Bugey, CCFR, and NOMAD.
A search for \nm $\to$ \ne oscillations has also been conducted by the
LSND collaboration.  Using \nm from $\pi^+$ decay in flight, the \ne
appearance is detected via the charged-current reaction
$C(\ne,e^-)X$. Two independent analyses are consistent with the above
signature, after taking into account the events expected from the \ne
contamination in the beam and the beam-off background.  If interpreted
as an oscillation signal, the observed oscillation probability of $2.6
\pm 1.0 \pm 0.5 \times 10^{-3}$, consistent with the evidence for 
oscillation in the \bnm $\to$ \bne channel described above. 
Fig.~\ref{miniboone} compares the
LSND region with the expected sensitivity from MiniBooNE, which was
recently approved to run at Fermilab~\cite{louis,Louis:1998qf}.
A possible confirmation of the LSND anomaly would be a discovery of
far-reaching implications.

\subsubsection{ Dark Matter}

Galaxies as well as the large scale structure in the Universe should
arise from the gravitational collapse of fluctuations in the expanding
universe. They are sensitive to the nature of the cosmological dark
matter. The data on cosmic background temperature anisotropies on
large scales performed by the COBE satellite~\cite{cobe} combined with
cluster-cluster correlation data e.g. from IRAS~\cite{iras} can not be
reconciled with the simplest COBE-normalized $\Omega_m=1$ cold dark
matter (CDM) model, since it leads to too much power on small
scales. Adding to CDM neutrinos with mass of few eV (a scale similar
to the one indicated by the LSND experiment
\cite{LSND}) corresponding to $\Omega_\nu \approx 0.2$, results in an
improved fit to data on the nearby galaxy and cluster
distribution~\cite{cobe2}.  The resulting Cold + Hot Dark Matter
(CHDM) cosmological model is the most successful $\Omega_m=1$ model
for structure formation, preferred by inflation.  However, other
recent data have begun to indicate a lower value for $\Omega_m$, thus
weakening the cosmological evidence favouring neutrino mass of a few
eV in flat models with cosmological constant $\Omega_\Lambda = 1 -
\Omega_m$~\cite{cobe2}.  Future sky maps of the cosmic microwave 
background radiation (CMBR) with high precision at the MAP and PLANCK
missions should bring more light into the nature of the dark matter
and the possible r\^ole of neutrinos \cite{Raffelt:1999zg}.  Another
possibility is to consider unstable dark matter scenarios
\cite{Gelmini:1984pe}. For example, an MeV range tau neutrino may
provide a viable unstable dark matter scenario \cite{ma1} if the \nt
decays before the matter dominance epoch. Its decay products would add
energy to the radiation, thereby delaying the time at which the matter
and radiation contributions to the energy density of the universe
become equal. Such delay would allow one to reduce the density
fluctuations on the smaller scales purely within the standard cold
dark matter scenario.  Upcoming MAP and PLANCK missions may
place limits on neutrino stability \cite{Hannestad:1999xy} and rule
out such schemes.

\subsubsection{  Pulsar Velocities}

One of the most challenging problems in modern astrophysics is to find
a consistent explanation for the high velocity of pulsars.
Observations \cite{veloc} show that these velocities range from zero
up to 900 km/s with a mean value of $450 \pm 50$ km/s.  An attractive
possibility is that pulsar motion arises from an asymmetric neutrino
emission during the supernova explosion. In fact, neutrinos carry more
than $99 \%$ of the new-born proto-neutron star's gravitational
binding energy so that even a $1 \%$ asymmetry in the neutrino
emission could generate the observed pulsar velocities.  This could in
principle arise from the interplay between the parity violation
present in weak interactions with the strong magnetic fields which are
expected during a SN explosion~\cite{Chugai,others}. However, it has
recently been noted~\cite{vilenkin98} that no asymmetry in neutrino
emission can be generated in thermal equilibrium, even in the presence
of parity violation. This suggests that an alternative mechanism is at
work.
Several neutrino conversion mechanisms in matter have been invoked as
a possible engine for powering pulsar motion.  They all rely on the
{\sl polarization} \cite{NSSV} of the SN medium induced by the strong
magnetic fields $10^{15}$ Gauss present during a SN explosion. This
would affect neutrino propagation properties giving rise to an angular
dependence of the matter-induced neutrino potentials. This would lead
in turn to a deformation of the "neutrino-sphere" for, say, tau
neutrinos and thus to an anisotropic neutrino emission.  As a
consequence, in the presence of non-vanishing $\nu_\tau$ mass and
mixing the resonance sphere for the $\nu_e-\nu_\tau$ conversions is
distorted.  If the resonance surface lies between the $\nu_\tau$ and
$\nu_e$ neutrino spheres, such a distortion would induce a temperature
anisotropy in the flux of the escaping tau-neutrinos produced by the
conversions, hence a recoil kick of the proto-neutron star.
This mechanism was realized in ref.~\cite{KusSeg96} invoking MSW
conversions \cite{MSW} with \mnt $\gsim$ 100 eV or so, assuming a
negligible $\nu_e$ mass. This is necessary in order for the resonance
surface to be located between the two neutrino-spheres.  It should be
noted, however, that such requirement is at odds with cosmological
bounds on neutrinos masses unless the $\tau$-neutrino is unstable.
On the other hand in ref.~\cite{ALS} a realization was proposed in the
resonant spin-flavour precession scheme (RSFP) \cite{RSFP}.  The
magnetic field would not only affect the medium properties, but would
also induce the spin-flavour precession through its coupling to the
neutrino transition magnetic moment \cite{SFP}.

Perhaps the simplest suggestion was proposed in
ref.~\cite{Grasso:1998tt} where the required pulsar velocities would
arise from anisotropic neutrino emission induced by resonant
conversions of massless neutrinos (hence no magnetic moment).

Raffelt and Janka~\cite{Janka:1999kb} have argued, however, that the
asymmetric neutrino emission effect was overestimated, since
the temperature variation over the deformed neutrino-sphere is
not an adequate measure for the anisotropy of the neutrino
emission. This would invalidate all neutrino conversion mechanisms,
leaving the pulsar velocity problem without any known viable
solution. One potential way out would invoke conversions into sterile
neutrinos, since the conversions would take place deeper in the
star. However, it is too early to tell whether or not it works
\cite{nuno98}.

\section{Fitting the Puzzles Together}
\vskip .1cm
 
Physics beyond the Standard Model is required in order to explain
solar and atmospheric neutrino data. While neutrino oscillations
provide an excellent fit, alternative mechanisms are still
viable. Thus it is still too early to tell for sure whether neutrino
masses and angles are really being determined experimentally.  Here we
assume the standard neutrino oscillation interpretation of the
data. While it can easily be accommodated in theories of neutrino
mass, in general the angles involved are not predicted, in particular
the maximal mixing indicated by the atmospheric data. It is suggestive
to consider a theory with {\sl bi-maximal} mixing of
neutrinos~\cite{bimax98} if the solar neutrino data are explained in
terms of the just-so solution. This is not easy to reconcile in a
predictive quark-lepton {\sl unification} scheme that relates lepton
and quark mixing angles, since the latter are known to be small. For
recent attempts to reconcile solar and atmospheric data in unified
models with specific texture anzatze, see
ref.~\cite{Lola:1998xp,Altarelli:1998ns}.
The story gets more complicated if one wishes to account also for the
LSND anomaly and for the hot dark matter~\cite{ptv92,pv93,cm93}.  As
we have seen the atmospheric neutrino data requires $\Delta m^2_{atm}$
which is much larger than the scale $\Delta {m^2}_\odot$ which is
indicated by the solar neutrino data. This implies that with just the
three known neutrinos there is no room, unless some of the
experimental data are discarded.

\subsection{Almost Degenerate Neutrinos}
\vskip .1cm

The only possibility to fit solar, atmospheric and HDM scales in a
world with just the three known neutrinos is if all of them have
nearly the same mass \cite{cm93}, of about $\sim$ 1.5 eV or so in
order to provide the right amount of HDM \cite{cobe2} (all three
active neutrinos contribute to HDM). This can be arranged in the
unification approach discussed in sec. 2 using the $M_L$ term present
in general in seesaw models. With this in mind one can construct,
e.g. unified \10 seesaw models where all neutrinos lie at the above
HDM mass scale ($\sim$ 1.5 eV), due to a suitable horizontal symmetry,
while the parameters $\Delta {m^2}_\odot$ \& $\Delta {m^2}_{atm}$
appear as symmetry breaking effects. An interesting fact is that the
ratio $\Delta {m^2}_\odot \:/\:\Delta {m^2}_{atm}$ appears as
${m_c}^2/{m_t}^2$~\cite{DEG}. There is no room in this case to
accommodate the LSND anomaly. To what extent this solution is
theoretically natural has been discussed recently in
ref.~\cite{Casas:1999ac}.

\subsection{Four-Neutrino Models}
\vskip .1cm

The simplest way to incorporate the LSND scale is to invoke a fourth
neutrino. It must be \21 singlet ensuring that it does not affect the
invisible Z decay width, well-measured at LEP.  The sterile neutrino
\ns must also be light enough in order to participate in the
oscillations together with the three active neutrinos. The theoretical
challenges we have are:
\bi 
\item
to understand what keeps the sterile neutrino light, since the \21
gauge symmetry would allow it to have a large bare mass
\item
to account for the maximal neutrino mixing indicated by the
atmospheric data, and possibly by the solar
\item
to account from first principles for the scales $\Delta m^2_{atm}$,
$\Delta {m^2}_\odot$ and $\Delta m^2_{LSND/HDM}$
\ei
With this in mind we have formulated the simplest maximally symmetric
schemes, denoted as $(e\tau)(\mu~s)$~\cite{ptv92} and $(es)(\mu\tau)$
~\cite{pv93}, respectively. One should realize that a given scheme
(mainly  the structure of the leptonic charged current) may be realized
in more than one theoretical model. For example, an alternative to the
model in ~\cite{pv93} was suggested in ref.~\cite{cm93}. There have
been many attempts to derive the above phenomenological scenarios from
different theoretical assumptions, as has been discussed here
\cite{ptvlate,smir}.
 
Although many of the phenomenological features arise also in other
models, here I concentrate the discussion mainly on the theories
developed in ref.~\cite{ptv92,pv93}. These are characterized by
a very symmetric mass spectrum in which there are two ultra-light
neutrinos at the solar neutrino scale and two maximally mixed almost
degenerate eV-mass neutrinos (LSND/HDM scale), split by the
atmospheric neutrino scale~\cite{ptv92,pv93}. The HDM problem requires
the heaviest neutrinos at about 2 eV mass~\cite{pvhdm}.
These scales are generated radiatively due to the additional Higgs
bosons which are postulated, as follows: $\Delta m^2_{LSND/HDM}$
arises at one-loop, while $\Delta m^2_{atm}$ and $\Delta {m^2}_\odot$
are two-loop effects.  Since these models pre-dated the LSND results,
they naturally focussed on accounting for the HDM problem, rather than
LSND. However, in the meantime the evidence for hot dark matter has
weakened, whereas LSND came into play. In contrast to the HDM problem,
the LSND anomaly, if confirmed, would be a more convincing indication
for the existence of a fourth light neutrino species, considering that
the HDM may be accounted for in a three neutrino degenerate scenario.

\begin{figure}[t]
\centerline{\protect\hbox{\psfig{file=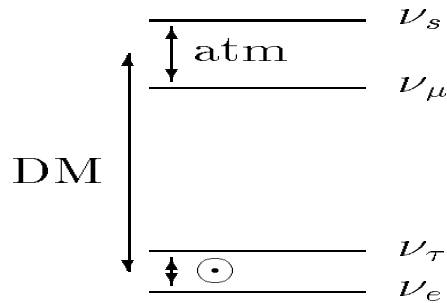,width=6cm,height=4cm}}}
\caption{$(e\tau)(\mu~s)$ scheme: \ne- \nt conversions explain the
solar neutrino data and \nm- \ns oscillations account for the
atmospheric deficit, ref.~\protect\cite{ptv92}.}
\label{ptv}
\vglue -.2cm
\end{figure}

\begin{figure}
\centerline{\protect\hbox{\psfig{file=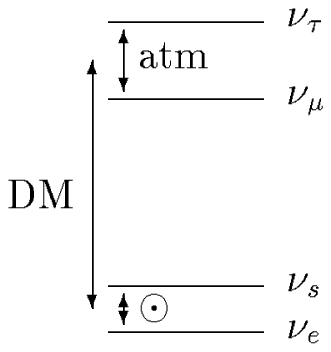,width=6cm,height=4cm}}}
\caption{$(es)(\mu\tau)$ scheme: \ne- \ns conversions explain the
solar neutrino data and \nm- \nt oscillations account for the
atmospheric deficit, ref.~\protect\cite{pv93}.}
\label{pv}
\vglue -.2cm
\end{figure}

The models in ~\cite{ptv92,pv93} are based only on weak-scale physics.
They {\sl explain} the lightness of the sterile neutrino, the large
lepton mixing required by the atmospheric neutrino data, as well as
the generation of the mass splittings responsible for solar and
atmospheric neutrino conversions as natural consequences of the
underlying lepton-number-like symmetry and its breaking.  They are
minimal in the sense that they add a single \21 singlet lepton to the
SM.  Before breaking the symmetry the heaviest neutrinos are exactly
degenerate, while the other two are still massless
\cite{OLDsterilemodel}.  After the global U(1) lepton symmetry breaks
the heavier neutrinos split and the lighter ones get mass.  The models
differ according to whether the \ns lies at the dark matter scale or
at the solar neutrino scale. In the $(e\tau)(\mu~s)$ scheme the \ns
lies at the LSND/HDM scale, as illustrated in \fig{ptv}
while in the alternative $(es)(\mu\tau)$ model, \ns is at the solar
\neu scale as shown in \fig{pv} \cite{pv93}.
In the $(e\tau)(\mu~s)$ case the atmospheric neutrino puzzle is explained
by \nm to \ns oscillations, while in $(es)(\mu\tau)$ it is due to
 \nm to \nt oscillations. Correspondingly, the deficit of solar
\neus is explained in the first case by \ne to \nt conversions, while
in the second the relevant channel is \ne to $\nu_s$.  

The presence of additional weakly interacting light particles, such as
our light sterile neutrino, is constrained by BBN since the \ns
would enter into equilibrium with the active neutrinos in the early
Universe (and therefore would contribute to $N_\nu^{max}$) via
neutrino oscillations
\cite{bbnsterile}, unless
%
$\Delta m^2 sin^42\theta \lsim 3\times 10^{-6}~~eV^2$
%
Here $\Delta m^2$ denotes the mass-square difference of the active and
sterile species and $\theta$ is the vacuum mixing angle.  However,
systematic uncertainties in the BBN bounds still caution us not to
take them too literally. For example, it has been argued that present
observations of primordial Helium and deuterium abundances may allow
up to $N_\nu = 4.5$ neutrino species if the baryon to photon ratio is
small~\cite{sarkar}. Adopting this as a limit, clearly both models
described above are consistent. Should the BBN constraints get tighter
\cite{Fiorentini:1998fv} e.g. $N_\nu^{max} < 3.5$ they could rule out
the $(e\tau)(\mu~s)$ model, and leave out only the competing scheme as
a viable alternative. However the possible r\^ole of a primordial
lepton asymmetry might invalidate this conclusion, for recent work on
this see ref.~\cite{Foot:1997qc}.

The two models would be distinguishable both at future solar as well
as atmospheric neutrino data. For example they may be tested in the
SNO experiment~\cite{SNO} once they measure the solar neutrino flux
($\Phi^{NC}_{\nu}$) in their neutral current data and compare it with
the corresponding CC value ($\Phi^{CC}_{\nu}$). If the solar neutrinos
convert to active neutrinos, as in the $(e\tau)(\mu~s)$ model, then
one expects $\Phi^{CC}_{\nu}/\Phi^{NC}_{\nu}$ around 0.5, whereas in
the $(es)(\mu\tau)$ scheme (\ne conversion to \ns), the above ratio
would be nearly $ \simeq 1$.  Looking at pion production via the
neutral current reaction $\nu_{\tau} + N \to \nu_{\tau} +\pi^0 +N$ in
atmospheric data might also help in distinguishing between these two
possibilities~\cite{vissani}, since this reaction is absent in the
case of sterile neutrinos, but would exist in the $(es)(\mu\tau)$
scheme.

If light sterile neutrinos indeed exist one can show that they might
contribute to a cosmic hot dark matter component and to an increased
radiation content at the epoch of matter-radiation equality. These
effects leave their imprint in sky maps of the cosmic microwave
background radiation (CMBR) and may thus be detectable with the very
high precision measurements expected at the upcoming MAP and PLANCK
missions as noted recently in ref.~\cite{Raffelt:1999zg}.

\subsection{MeV Tau Neutrino}
\vskip .1cm

In ref.~\cite{JV95} a model was presented where an unstable MeV
Majorana tau neutrino naturally reconciles the cosmological
observations of large and small-scale density fluctuations with the
cold dark matter picture. The model assumes the spontaneous violation
of a global lepton number symmetry at the weak scale.  The breaking of
this symmetry generates the cosmologically required decay of the \nt
with lifetime $\tau_{\nu_\tau} \sim 10^2 - 10^4$ sec, as well as the
masses and oscillations of the three light \neus \ne, \nm and $\nu_s$
which may account for the present solar and atmospheric data, though
this will have to be checked.  One can also verify that the BBN
constraints can be satisfied.

\section{In conclusion}
\vskip .1cm

The confirmation of an angle-dependent atmospheric neutrino deficit
provides, together with the solar neutrino data, a strong evidence for
physics beyond the Standard Model. Small neutrino masses
provide the simplest, but not unique, explanation of the data.  If the
LSND result stands the test of time, this would be a puzzling
indication for the existence of a light sterile neutrino.
The two most attractive schemes to reconcile underground observations
with LSND invoke either \ne- \nt conversions to explain the solar
data, with \nm- \ns oscillations accounting for the atmospheric
deficit, or the opposite. These two basic schemes have
distinct implications at future solar \& atmospheric neutrino
experiments. SNO and Super-Kamiokande have the potential to
distinguish them due to their neutral current sensitivity.

Allowing for alternative explanations of the data from underground
experiments one can still live with massless non-standard neutrinos or
even very heavy neutrinos, which may naturally arise in many
models. Although cosmological bounds are a fundamental tool to
restrict neutrino masses, in many theories heavy neutrinos will either
decay or annihilate very fast, thereby loosening the cosmological
bounds. From this point of view, {\sl neutrinos can have any mass
presently allowed by laboratory experiments}, and it is therefore
important to search for manifestations of heavy neutrinos at the
laboratory in an unbiased way.

Last but not least, though most of the recent excitement comes from
underground experiments, one should note that models of neutrino mass
may lead to a plethora of new signatures which may be accessible also
at accelerators, thus illustrating the complementarity between the two
approaches in unravelling the properties of neutrinos and probing for
signals beyond the Standard Model~\cite{desert}.

\vskip .3cm

I am grateful to the Organizers for the kind hospitality at
Corfu. This work was supported by DGICYT grant PB95-1077 and by the
EEC under the TMR contract ERBFMRX-CT96-0090.

\end{document}